\documentclass[aps,twocolumn,showpacs]{revtex4}
\usepackage{pstricks}
\usepackage{tabularx}
\newcolumntype{Z}{>{\centering\arraybackslash}X}
\usepackage{graphicx}
\usepackage{amsmath}
\usepackage{amssymb}
\begin{document}
\title{Controlling quantum systems by embedded dynamical decoupling schemes}
\author{O.~Kern and G.~Alber}
\affiliation{Institut f\"ur Angewandte Physik, Technische Universit\"at Darmstadt, 64289 Darmstadt, Germany}
\date{Received: date / Revised version: date}
\begin{abstract}
A dynamical decoupling method is presented which is based on embedding a deterministic decoupling scheme into 
a stochastic one. This way it is possible to combine the advantages of both methods and to 
increase the suppression of undesired perturbations of quantum systems significantly even for long interaction times.
As a first application the stabilisation of a quantum memory is discussed which is perturbed by
one-and two-qubit interactions.
\end{abstract}

\maketitle
The desire to control quantum systems 
and to push characteristic quantum phenomena
more and more into the macroscopic domain necessitates the development of suitable methods 
which are capable of cancelling undesired perturbations as much as possible.
Such methods are not only of significant interest
for the further development of quantum information science but are also important in more traditional
areas of quantum physics, such as high-resolution spectroscopy.
So far, two different strategies have been pursued to suppress undesired perturbations in quantum systems.
First of all,
one may cancel such perturbations by appropriate 
sequences of control measurements and unitary recovery operations. This approach is at the heart of quantum error correction \cite{error}.
It requires the use of redundant quantum states
and, in principle, it is capable of correcting errors perfectly.
However, with nowaday´s experimental techniques it is difficult to realize typical 
syndrome measurements and recovery operations required.

The strategy pursued in dynamical decoupling methods \cite{dec1,dec2,dec3,dec4,dec5,dec6,Roetteler,bang-bang,Knill,Kern} is different. 
Thereby,
one uses controlled unitary dynamics to suppress
undesired perturbations as much as possible. As these methods are not based on redundancy and quantum measurements
they tend to be implementable in realistic scenarios more easily.
Depending on 
the type of dynamics applied
one distinguishes
deterministic and stochastic decoupling methods.
Deterministic methods \cite{dec1,dec2,dec3,dec4,dec5,dec6,Roetteler,bang-bang} 
are already well developed.
Numerous deterministic sequences of unitary operations are already known which are able to suppress  
few-qubit perturbations.
Stochastic decoupling methods \cite{Knill,Kern} have been introduced only recently.
In particular, it was demonstrated that for sufficiently
long interaction times  stochastic
decoupling methods offer advantages over deterministic ones. Typically, they lead to a linear-in-time fidelity decay whereas the fidelity decay
of deterministic methods is  quadratic in time. Nevertheless, for sufficiently short interaction times
the error suppression capabilities of deterministic methods are typically stronger than of stochastic ones.

In this letter
it is demonstrated that the advantages of both methods can be combined 
by embedding  an 
appropriate deterministic decoupling scheme into a stochastic one.
As a result, one achieves the significant suppression 
of perturbations which is characteristic for deterministic decoupling schemes and, in addition,
one also produces a rather slow linear-in-time fidelity decay which is typical for stochastic decoupling methods.
Motivated by the potential impact these error suppression methods might have on the future development of
quantum information science
in the following
the basic ideas of these embedded
decoupling methods are demonstrated for a quantum memory which is perturbed by inter-qubit couplings.

\begin{figure}
\small
\begin{center}
\begin{pspicture}(0,0)(\columnwidth,-5.6)
%
\psline[linewidth=1pt,arrowsize=7pt]{<-}(0.4,-1.)(8.5,-1.)
\rput[t](0.25,-0.5){\small (a)}
\rput[t](0.3,-1.25){\footnotesize time}
\psline(8.5,-1.1)(8.5,-0.9)\rput[t](8.5,-1.3){\small $t_0$}\rput[b](8.5,-0.8){\small $\hat{d}_0$}
\psline(7.0,-1.1)(7.0,-0.9)\rput[t](7.0,-1.3){\small $t_1$}\rput[b](7.0,-0.8){\small $\hat{d}_1 \hat{d}_0^\dagger$}
\psline(5.5,-1.1)(5.5,-0.9)\rput[t](5.5,-1.3){\small $t_2$}\rput[b](5.5,-0.8){\small $\hat{d}_2 \hat{d}_1^\dagger$}
\psline[linewidth=1pt,linestyle=dashed,linecolor=white](3.5,-1.)(4.5 ,-1.)
\psline(2.5,-1.1)(2.5,-0.9)\rput[t](2.5,-1.3){\small $t_N\equiv T_c$}\rput[b](2.5,-0.8){\small $\hat{d}_0 \hat{d}_{N-1}^\dagger$}
\psline(1.0,-1.1)(1.0,-0.9)\rput[t](1.0,-1.3){\small $t_{N+1}$}\rput[b](1.0,-0.8){\small $\hat{d}_1 \hat{d}_0^\dagger$}
\psline[linewidth=1pt,arrowsize=7pt]{<-}(0.5,-3.)(8.5,-3.)
\rput[t](0.25,-2.5){\small (b)}
\rput[t](0.3,-3.25){\footnotesize time}
\psline(8.5,-3.1)(8.5,-2.9)\rput[t](8.5,-3.3){\small $0$}\rput[b](8.5,-2.8){\small $\hat{r}_0$}
\psline(7.0,-3.1)(7.0,-2.9)\rput[t](7.0,-3.3){\small $\Delta t$}\rput[b](7.0,-2.8){\small $\hat{r}_1 \hat{r}_0^\dagger$}
\psline(5.5,-3.1)(5.5,-2.9)\rput[t](5.5,-3.3){\small $2\Delta t$}\rput[b](5.5,-2.8){\small $\hat{r}_2 \hat{r}_1^\dagger$}
\psline(4.0,-3.1)(4.0,-2.9)\rput[t](4.0,-3.3){\small $3\Delta t$}\rput[b](4.0,-2.8){\small $\hat{r}_3 \hat{r}_2^\dagger$}
\psline[linewidth=1pt,linestyle=dashed,linecolor=white](2.0,-3.)(3.0 ,-3.)
\psline[linewidth=1pt,arrowsize=7pt]{<-}(0.5,-5.)(8.5,-5.)
\rput[t](0.25,-4.5){\small (c)}
\rput[t](0.3,-5.25){\footnotesize time}
\psline(8.5,-5.1)(8.5,-4.9)\rput[t](8.5,-5.3){\small $t_0$}\rput[b](8.5,-4.8){\small $\hat{d}_0 \hat{r}_0$}
\psline(7.0,-5.1)(7.0,-4.9)\rput[t](7.0,-5.3){\small $t_1$}\rput[b](7.0,-4.8){\small $\hat{d}_1 \hat{d}_0^\dagger$}
\psline(5.5,-5.1)(5.5,-4.9)\rput[t](5.5,-5.3){\small $t_2$}\rput[b](5.5,-4.8){\small $\hat{d}_2 \hat{d}_1^\dagger$}
\psline[linewidth=1pt,linestyle=dashed,linecolor=white](3.5,-5.)(4.5 ,-5.)
\psline(2.5,-5.1)(2.5,-4.9)\rput[t](2.5,-5.3){\small $t_N\equiv T_c$}\rput[b](2.5,-4.8){\small $\hat{d}_0 \hat{r}_1 \hat{r}_0^\dagger \hat{d}_{N-1}^\dagger$}
\psline(1.0,-5.1)(1.0,-4.9)\rput[t](1.0,-5.3){\small $t_{N+1}$}\rput[b](1.0,-4.8){\small $\hat{d}_1 \hat{d}_0^\dagger$}
\end{pspicture}
\end{center}
\normalsize
\caption{Schematic representation of the deterministic bang-bang method (a), the PAREC-method (b), and an associated
embedded decoupling scheme:
The time evolution between subsequent instantaneously applied unitary operations
is governed by the perturbing Hamiltonian $\hat{H}_0$.
In the embedded decoupling scheme (c) the deterministic decoupling cycle of (a)
is embedded between any two subsequent random unitary operations
of the PAREC-method (b).
\label{decoupling}}
\end{figure}
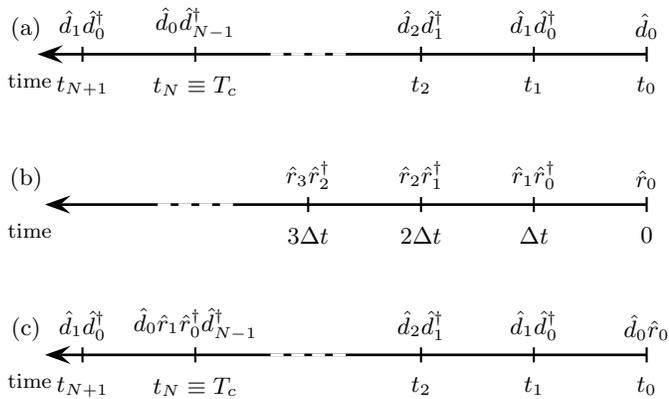

Let us start by summarising briefly the main ideas underlying deterministic
dynamical decoupling methods \cite{dec1,dec2,dec3,dec4,dec5,dec6,Roetteler,bang-bang} in more detail. For this purpose we consider
a quantum state $|\psi\rangle$
of a $n_q$-qubit quantum memory which is perturbed by an unknown inter-qubit coupling Hamiltonian, say $\hat{H}_0$.
The resulting coherent errors
may be suppressed 
by modifying the $n_q$-qubit dynamics with the help of an appropriate periodic deterministic control sequence of 
unitary operations. Within the framework of the bang-bang methods\cite{dec1}, for example, such a periodic
cycle of duration $T_c$ consists of $N$ unitary operations. They are of the form 
$\hat{d}_j \hat{d}^{\dagger}_{j-1}$ with properly chosen unitary operations $\hat{d}_j$ 
which are applied (approximately) instantaneously at times $t_0,...,t_{N-1}$ (compare with Fig.\ref{decoupling}a).
Thus, after one cycle the resulting dynamics of the quantum memory are described by the unitary time evolution operator \cite{Knill}
\begin{multline}
\hat{U}(T_c) = \hat{d}^{\dagger}_{N-1}e^{-i\hat{H}_0(t_{N}-t_{N-1})/\hbar} \hat{d}_{N-1}\hat{d}^{\dagger}_{N-2} \cdots \label{dynamics}\\
\cdots \hat{d}_1\hat{d}^{\dagger}_{0} e^{-i\hat{H}_0(t_{1}-t_0)/\hbar} \hat{d}^{\dagger}_{0}
\equiv \mathcal{T} \prod_{j=0}^{N-1}  e^{-i\hat{\tilde{H}}_j(t_{j+1}-t_j)/\hbar}
\end{multline}
with $t_0 \equiv 0$, $t_N \equiv T_c$, and 
with the `interaction-picture' Hamiltonians
$\hat{\tilde{H}}_j =
\hat{d}_{j}^{\dagger}
\hat{H}_0 
\hat{d}_j$.
In order to suppress the perturbing influence of $\hat{H}_0$ as much as possible this control cycle
should fulfil the requirement
$\sum_{j=0}^{N-1}  \hat{\tilde{H}}_j (t_{j+1}-t_j) = 0$.
This condition guarantees that 
at multiples of the cycle time
the perturbation is cancelled by the control sequence
at least to lowest order perturbation theory.
Nevertheless,
as this cancellation is not perfect, repeated applications of the deterministic
control cycle lead to a coherent accumulation of the residual perturbation characterised
by the Hamiltonian $\hat{\overline{H}}$ with
\begin{eqnarray}
\hat{{U}}(T_c) = 
e^{-i\hat{\overline{H}}T_c/\hbar}.
\label{res}
\end{eqnarray}
As a result any linear superposition of eigenstates of $\hat{\overline{H}}$ is dephased.
Thus, 
the short-time evolution of the
fidelity $f(T)$ of the quantum 
state $|\psi\rangle$ is given by \cite{Knill}
\begin{eqnarray}\label{fidelity}
f(T) &\equiv& \vert \langle \psi | \hat{U}(T=nT_c)|\psi\rangle \vert^2 = \nonumber\\
&& 1 - (T\Delta \overline{H}/\hbar)^2 + \mathcal{O} \left((T\Delta \overline{H}/\hbar)^3 \right)
\end{eqnarray}
with
the relevant energy uncertainty
of the residual interaction 
$(\Delta \overline{H})^2 = \langle \psi |(\hat{\overline{H}}- \langle \psi  |\hat{\overline{H}}|\psi\rangle)^2|\psi\rangle$.
An upper bound of the residual interaction is given by the norm of its Hamiltonian
$\Vert \hat{\overline{H}} \Vert$.
Using Eq.\eqref{res} and basic properties of the logarithmic function we may obtain the inequalities
\begin{eqnarray}
\Delta \hat{\overline{H}} &\leq& 
\Vert \hat{\overline{H}} \Vert \leq 
-\frac{\ln\left( 
2 - e^{\Vert \hat{{H}}_0 \Vert T_c/\hbar} +
\frac{ \Vert \hat{{H}}_0 \Vert T_c}{\hbar}
\right)}{T_c/\hbar}.
\label{hbar}
\end{eqnarray}
Thus, for small cycle times $T_c$ with $\Vert \hat{{H}}_0 \Vert T_c/\hbar \ll 1$
a lower bound of the fidelity 
is given by \cite{Knill,comment}
\begin{eqnarray}
f(T) \geq 1 - (\frac{\Vert \hat{H}_0 \Vert^2 T T_c}{2\hbar^2})^2.
\label{fid3}
\end{eqnarray}
In general,
deterministic decoupling schemes suffer from two major drawbacks \cite{Knill}. Firstly, some perturbations may require rather
long control cycles so that
a significant error suppression cannot be achieved.  Secondly, it is unclear how to suppress perturbations 
which change on time scales short in comparison with cycle times
required.

In order to overcome these drawbacks stochastic decoupling methods have been proposed recently \cite{Knill,Kern}.
In the Pauli-random-error-correction-(PAREC)-method \cite{Kern}, for example, the periodically repeated
control cycles of deterministic methods are replaced by a single random sequence
of (approximately) instantaneous uncorrelated unitary operations.
Thus, the deterministic unitary operations
$\hat{d}_j$ of the bang-bang methods are replaced by unitary transformations $\hat{r}_j$ 
each of which is a product of uncorrelated random one-qubit Pauli-operators,
i.\,e. $\hat{r}_j \in \{ \hat{\mathbf{1}}, \hat{X}, \hat{Y}, \hat{Z} \}^{\otimes n_q}$,
with the Pauli spin operators $\hat{X}$, $\hat{Y}$, and $\hat{Z}$.
Within a time interval of duration $T$
the resulting dynamics 
are described by a unitary time evolution operator of the form of Eq. \eqref{dynamics} with
the replacement $T_c\to T$
and with the `interaction-picture' Hamiltonians $\hat{\tilde{H}}_j\to \hat{r}_j^{\dagger} \hat{H}_0 \hat{r}_j$.
In order to suppress the perturbing influence of $\hat{H}_0$ 
the random sequence of unitary Pauli operations $\hat{r}_j$ should be statistically independent.
Provided these uncorrelated random Pauli-operations are applied after equal time intervals of duration $\Delta t$
a lower bound of the mean fidelity is given by
\cite{Knill,Kern,Shepelyansky}
\begin{equation}
\mathbb{E} f(T) = 1 - \Gamma T  + \mathcal{O} \left( (\Gamma T)^2\right) \geq 1 -
\frac{\Vert \hat{H}_0 \Vert^2 T\Delta t}{\hbar^2}, \label{fid2}
\end{equation}
if the interaction time $T\equiv n\Delta t$ is so small that $\mid1 - {\mathbb{E}}f(T)\mid \ll 1$.
($\mathbb{E}$ denotes the statistical averaging over the uncorrelated random unitary operations $\hat{r}_j$.)
Consistent with Fermi's Golden rule 
the decay rate is given by
$\Gamma = (2\pi/\hbar)[\Delta t/(2\pi\hbar)]\Delta H_0^2$
and 
$(\Delta {H}_0)^2 = \langle \psi |(\hat{{H}}_0- \langle \psi  |\hat{{H}}_0|\psi\rangle)^2|\psi\rangle$
is the relevant energy uncertainty of the original perturbation.
The lower bound of \eqref{fid2} \cite{Knill,comment} is based on the
inequality
$(\Delta {H}_0)^2 \leq \Vert \hat{H}_0 \Vert^2$.

According to relation \eqref{fid2}
stochastic decoupling methods, such as the PAREC-method, offer the advantage
that residual errors which are not corrected by the randomly applied unitary operations do not
add coherently so that the fidelity decays linearly with interaction time $T$
and not quadratically as in the case of deterministic methods.
As a consequence, 
for sufficiently long interaction times 
stochastic decoupling methods offer a higher degree of error suppression than deterministic ones.
But, for sufficiently short interaction times  deterministic decoupling
methods are still expected to be more accurate because the lower bounds of their fidelities scale with
$\Vert \hat{H}_0 \Vert^4$ (compare with \eqref{fid3}) in contrast to the $\Vert \hat{H}_0 \Vert^2$-scaling predicted by relation
\eqref{fid2}.

By embedding a deterministic decoupling scheme into a stochastic one
errors are suppressed to a large extent not only for
short but also for long interaction times without requiring any major additional effort. The basic idea of such an embedding
scheme
based on the recently proposed
PAREC-method is depicted in Fig.\ref{decoupling}c.
In order to
stabilise a quantum state over a time interval of magnitude $T$
we first of all take advantage of the PAREC-method by applying (approximately) instantaneous
random Pauli operations after each time interval of duration $\Delta t$.
In addition, in order to get rid of
large parts of the undesired perturbation $\hat{H}_0$ within each of these time intervals
we embed
a deterministic decoupling scheme with cycle time $T_c = \Delta t$
between any two of these random Pauli operations. As a result,
within each time interval of duration $\Delta t$ 
the time evolution
is now governed by 
the residual interaction $\hat{\overline{H}}$. 
Furthermore, large parts of it
are suppressed by the randomisation of the PAREC-method.
Thus, this embedding procedure decreases the fidelity decays of Eqs.\eqref{fidelity} and \eqref{fid2} significantly.
In particular, using the replacements $\hat{H}_0~\to~\hat{\overline{H}}$,
$\Delta t \to T_c$ in the lower bound of Eq.\eqref{fid2}, and invoking relation \eqref{hbar}
for $\Vert \hat{H}_0 \Vert T_c/\hbar \ll 1$
one obtains the lower bound
\begin{equation}
\mathbb{E} f(T=n T_c) \geq 1 - \left( \frac{\Vert \hat{{H}}_0 \Vert^2 T_c}{2\hbar^2} \right)^2 T T_c
\label{fidemb}
\end{equation}
for the mean fidelity of the resulting quantum state. 
It
indicates already a significantly slower fidelity decay than the corresponding decays of the deterministic and stochastic procedures of relations \eqref{fid3} and \eqref{fid2}.
In particular, this lower bound of the fidelity decays linearly with $T$ and, in addition, it scales with
$\Vert \hat{H}_0 \Vert^4$.

\begin{figure}
\centerline{\includegraphics[width=.95\columnwidth]{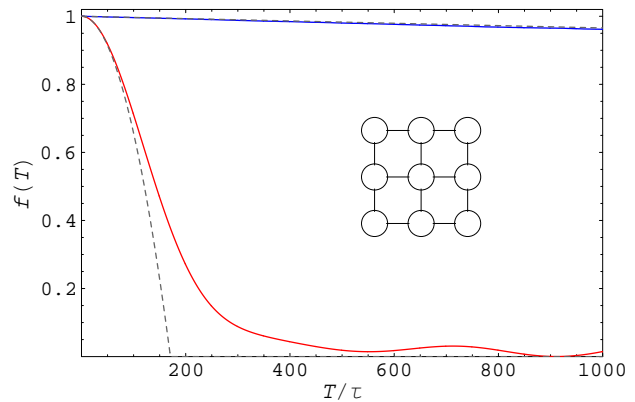}}
\caption{Time evolution of the fidelity $f(T)$:
without error correction (full lower curve),  
resulting from the PAREC-method (full upper curve). The dashed curves indicate the approximations of
\eqref{fid2}
and \eqref{fid4}.
The interaction time $T$ is plotted in units of the time interval $\tau $ between subsequent randomly applied uncorrelated Pauli-operations.
The rectangular array of the nine qubits  constituting the quantum memory is depicted in the inset.
The lines indicate the
non-zero values of the coupling parameters  $J_{kl}\tau$. They are chosen randomly  from the interval
$[-\sqrt{3}\times 10^{-3},\sqrt{3}\times 10^{-3}]$ (static imperfection).
\label{qubits}}
\end{figure}

Let us now investigate this fidelity decay quantitatively in more detail.
For this purpose we consider a quantum memory consisting of $n_q = 9$ distinguishable
qubits which are arranged in an equidistant two-dimensional
array (compare with Fig.\ref{qubits}).
This quantum memory is assumed to be perturbed by Ising- and Heisenberg-type qubit interactions of the form
$\hat{H}_0/\hbar = \sum_{k=0}^{n_q-1} \delta_k \hat{Z}_k + \sum_{k<l=0}^{n_q-1} J_{kl}(\hat{X}_k\hat{X}_l +\hat{Y}_k\hat{Y}_l +\hat{Z}_k\hat{Z}_l)$
with the time independent perturbation parameters $\delta_k$ and $J_{kl}$.
Initially, the quantum memory is assumed to be prepared in the coherent state
$|\psi\rangle = \left({2}/{D}\right)^{1/4}
\sum_{m=0}^{D-1} e^{-i\pi m  -\pi (m - D/2)^2/D} |m\rangle$
with the dimension of the Hilbert space $D=2^{n_q}$ and
with $| m \equiv \sum_{l=0}^{D-1} 2^l i_l \rangle = |i_1 \rangle\otimes   |i_2 \rangle\otimes   \cdots |i_{n_q} \rangle$ ($i_l\in\{0,1\}$) denoting
the $D$ orthonormal basis states of the computational basis.

The resulting time evolution of the fidelity 
is depicted  by the full lower curve of Fig.\ref{qubits}.
For sufficiently short interaction times $T$ the fidelity decay is given by 
\begin{eqnarray}
f(T) &=& 
  1 - (T\Delta H_0/\hbar)^2 + {\mathcal{O}}((T\Delta H_0/\hbar)^3).
\label{fid4}
\end{eqnarray}
The resulting quadratic-in-time decay reflects the dephasing of the initially prepared quantum state $|\psi\rangle$.
At larger interaction times the dynamics are governed by recurrence- and revival-phenomena which characterise the
intricate quantum interferences involved \cite{wavepackets}.
The extent to which the PAREC-method is capable of stabilising the initially prepared quantum state
is exemplified by the full upper curve of Fig.\ref{qubits}.
{The fidelity was obtained by averaging over 200 numerical runs.}
As expected, the resulting mean fidelity decay is linear in time and  is well described by the
decay rate $\Gamma$ of relation \eqref{fid2} (dashed upper curve of Fig.\ref{qubits}).

\begin{figure}
\centerline{\includegraphics[width=.95\columnwidth]{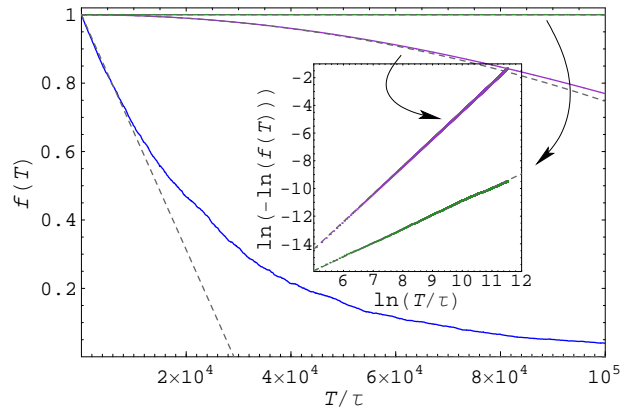}}
\caption{Fidelity decays resulting from the deterministic bang-bang method (full middle curve),
the PAREC-method (full lower curve),
and from the associated embedded decoupling scheme (full upper curve):
The corresponding approximate fidelities are indicated by dashed lines.
The interaction time $T$ is plotted in units of the time interval $\tau$ between subsequently applied unitary operations.
The inset shows the corresponding double-logarithmic plot.
\label{fidcomp}}
\end{figure}

As the perturbing interaction $\hat{H}_0$ involves one-and two-qubit operations only an efficient deterministic decoupling cycle can be realized by $32$ deterministic unitary operations $\hat{d}_j$
{applied at equidistant time steps $\tau$}
within the framework of a bang-bang decoupling scheme \cite{Roetteler,bang-bang}.
These operations are products of Pauli-spin operators of the form $\hat{d} = \sigma_{l_1}\otimes \cdots\otimes \sigma_{l_{n_q}}$
with $l_k \in \{0,1,2,3\}$ and with $\hat{\sigma}_0 = \hat{\mathbf{1}}, \hat{\sigma}_{1} = \hat{X}, \hat{\sigma}_{2} = \hat{Y}, \hat{\sigma}_{3} = \hat{Z}$.
The $32$ relevant values $(l_1,...,l_{n_q})$  are given by the orthogonal array $OA(32,9,4,2)$ of Ref.\cite{detscheme}.
The resulting time evolution of the fidelity is depicted by the full middle curve of Fig.\ref{fidcomp}.
It is apparent that 
this deterministic decoupling method leads to a significant slowing down of the  fidelity decay in comparison with the uncorrected time evolution
(full lower curve of Fig.\ref{qubits}).
Furthermore, for the interaction times considered this deterministic decoupling procedure also achieves a higher degree of stabilisation than the PAREC-method (lowest full curve of Fig.\ref{fidcomp}).
This may be traced back to the rather small residual interaction.
Furthermore, this fidelity decay is well approximated by relation \eqref{fidelity}.
Nevertheless, in view of its quadratic-in-time decay for sufficiently long interaction times $T$ this deterministic decoupling scheme will eventually become less effective than the linear-in-time decaying PAREC-method.

The fidelity decay originating from the embedded decoupling scheme in which this deterministic scheme is embedded into the PAREC-method is exemplified by the full uppermost curve of Fig.\ref{fidcomp}.
Again the fidelity was obtained by averaging over 200 numerical runs.
Over the whole range of interaction times the resulting fidelity is significantly higher than the corresponding values of both
the deterministic and the stochastic decoupling method.
Thereby, the lower bound of inequality \eqref{fidemb} strongly underestimates the actually achievable degree of stabilisation.
A much better approximation of the relevant fidelity decay is obtained by evaluating the rate $\Gamma$ of Eq.\eqref{fid2} on the basis of the
residual interaction $\hat{\overline{H}}$.
The slow linear-in-time exponential decay of the embedded scheme and the more rapid quadratic-in-time exponential decay of the deterministic scheme are also apparent from the inset of Fig.\ref{fidcomp}.

In summary, a generalised decoupling method has been presented which is based on the embedding of a deterministic scheme into a stochastic one.
Thus it is possible to combine the 
strong error suppression  properties of deterministic methods at sufficiently short interaction times
with the particularly advantageous properties of stochastic decoupling methods at longer interaction times.
The example presented demonstrates the high degree of error suppression achievable by such embedded decoupling schemes.
Though this example has focused on the stabilisation of a quantum memory's quantum state,
it is expected that
embedded dynamical decoupling schemes also offer new and promising perspectives
for stabilising the dynamical evolution of quantum systems or of quantum algorithms
against undesired perturbations at least in cases in which the relevant deterministic method is able to achieve this goal.

This work is supported by the 
EU IST-FET project EDIQIP and by the DFG (SPP-QIV).


\begin{thebibliography}{10}
\bibitem{comment} Rigorous lower bounds of the fidelities of Eqs.\eqref{fid3} and \eqref{fid2} valid for arbitrary
interaction times $T$ are also available \cite{Knill}.
Typically they overestimate the fidelity decay significantly.
\end{thebibliography}
\end{document}